\documentclass[intlimits,twoside,a4paper]{article}

\usepackage{amsmath,amssymb}
\usepackage{graphicx}

\usepackage[T2A]{fontenc}
\usepackage[cp1251]{inputenc}

\usepackage[eqsecnum]{cmpj2}

\issue{2014}{17}{3}{33006}
\doinumber{10.5488/CMP.17.33006}

\title[Jamming and percolation of parallel squares]{Jamming and percolation of parallel squares in single-cluster growth
model\thanks{This work is dedicated to the memory of Professor Alexander I. Olemskoi.}}

\author[I.A.~Kriuchevskyi  \textsl{et al.}]{
I.A.~Kriuchevskyi\refaddr{label1}, 
 L.A.~Bulavin\refaddr{label1}, 
 Yu.Yu.~Tarasevich\refaddr{label2}, 
 N.I.~Lebovka\refaddr{label3}\thanks{E-mail: lebovka@gmail.com}}

\addresses{
\addr{label1} Taras Shevchenko Kiev National University, Department of Physics,  2 Academician Glushkov Avenue, \\ 031127 Kyiv, Ukraine
\addr{label2} Astrakhan State University, 20a Tatishchev St., 414056 Astrakhan, Russia
\addr{label3} Institute of Biocolloidal Chemistry named after F.D. Ovcharenko of the National Academy of Sciences of Ukraine,
42 Academician Vernadsky Boulevard, 03142 Kiev, Ukraine
}

\authorcopyright{I.A.~Kriuchevskyi, L.A.~Bulavin,  Yu.Yu.~Tarasevich, N.I.~Lebovka, 2014}

\date{Received May 8, 2014}

\begin{document}

\maketitle
\begin{abstract}
This work studies the jamming and percolation of parallel squares in a single-cluster growth model. The Leath-Alexandrowicz method was used to grow a cluster from an active seed site. The sites of a square lattice were occupied by addition of the equal size $k \times k$ squares (E-problem) or a mixture of $k \times k$ and $m \times m$ ($m \leqslant  k$) squares (M-problem). The larger $k \times k$ squares were assumed to be active (conductive) and the smaller $m \times m$ squares were assumed to be blocked (non-conductive).  For equal size $k \times k$ squares (E-problem) the value of $p_j = 0.638 ± 0.001$ was obtained for the jamming concentration in the limit of $k\rightarrow\infty$. This value was noticeably larger than that previously reported for a random sequential adsorption model,  $p_j = 0.564 ± 0.002$. It was observed that the value of percolation threshold $p_{\mathrm{c}}$ (i.e., the ratio of the area of active $k \times k$ squares and the total area of $k \times k$ squares in the percolation point) increased with an increase of $k$. For mixture of $k \times k$ and $m \times m$ squares (M-problem), the value of $p_{\mathrm{c}}$ noticeably increased with an increase of $k$ at a fixed value of $m$ and approached  1 at $k\geqslant  10m$. This  reflects that percolation of larger active squares in M-problem can be effectively suppressed in the presence of smaller blocked squares.%


\end{abstract}

\section{\label{sec:introduction}Introduction}

Percolation and jamming problems for extended objects of various shapes and sizes, deposited on the lattices in two dimensions (2d), continuously attract great interest~\cite{Wang1998,Privman2000,Budinski2012}. The model of random sequential adsorption (RSA) is frequently used to form a random deposit on a substrate. In RSA model, the newly placed particle cannot overlap with the previously deposited ones, the adsorbed objects remain fixed and the final state is a disordered one (known as the jamming state)~\cite{Evans1993}. The fraction of the total surface, occupied by the adsorbed particles, is called the jamming concentration, $p_j$.  The objects with different shapes and sizes (e.g., linear~\cite{Manna1991,Becklehimer1992,Leroyer1994, Vandewalle2000,Kondrat2001,Lebovka2011, Tarasevich2012, Longone2012} and flexible (polymer-like)~\cite{Pawlowska2012,Pawlowska2013} $k$-mers (particles occupying $k$ adjacent sites), T-shaped objects and crosses~\cite{Adamczyk2008}, squares~\cite{Nakamura1987,Vigil1989, Dickman1991}, disks~\cite{Connelly2014}, ellipses~\cite{Evans1993}) have been studied, and data of these studies show that the value of $p_j$  strongly depends on the object shape and size.

The square-shaped  particles on planar substrates have been studied in numerous works as useful objects for a description of both fundamental and practical problems. The squares have been used as a model of anisotropic 2d ``molecules'' in equilibrium systems using different theoretical and Monte Carlo approaches~\cite{Zwanzig1956, Hoover1964, Carlier1972, Hoover2009}. The highest density for square particles is $1$, however, even in that state the particles can be arranged into uncorrelated parallel rows with non-crystalline order. Parallel squares appear to exhibit a second-order melting transition (with the pressure being continuous and the compressibility discontinuous) at a concentration $p$ of $0.79$~\cite{Hoover2009}. Square-shaped  particles are also interesting as potential mesogens. Monte Carlo simulations of squares have found a tetratic intermediate phase with the quasi-long-range orientational order and the translational order decaying faster than algebraically~\cite{Wojciechowski2004}. The rapid change of the orientational order of squares was observed at coverage of $\approx 0.69$.

Different variants of a non-equilibrium RSA model for squares deposited on 2d substrates were also developed. The coordination RSA model for $k \times k$ squares (squares of side $k$) deposited on a square lattice was studied numerically~\cite{Rodgers1993}. In this model, the squares are not allowed to touch one another if the number of contact exceeds the predefined value of average coordination number $c$. It was shown that the jamming coverage is $\approx 0.56$ in the limit of $k\rightarrow\infty$, independent of $c$.  Moreover, for an average coordination of about $2.4$, the jamming coverage was $\approx0.58$, independent of the size $k$. The effect of size distribution on the jamming coverage of parallel squares on the substrate was recently studied~\cite{Vieira2011}. A power distribution of square sizes can lead to a much larger packing density than for the equal size squares.

The numerical simulations of RSA deposition of $k \times k$ squares on a square lattice gave the following power dependence of jamming concentration $p_j(k)$~\cite{Nakamura1986,Nakamura1986a}:
\begin{equation}\label{eq:Nakamura}
|p_j -p_j^\infty|\propto k^{-\alpha},
\end{equation}
where  $p_j^\infty=0.564\pm 0.002$ corresponds to the continuous limit ($k\rightarrow\infty$) and  $\alpha\simeq 1$. Thus, the jamming concentration is in inverse proportion to the size of the square, $k$.

However, the percolation for the RSA model of $k \times k$ squares was observed only
at small values of $k$ ($k=1-3$). However, only finite clusters
of $k \times k$ squares were observed at saturation coverage for $k\geqslant  4$.
The values of jamming $p_j$ and percolation $p_{\mathrm{c}}$
concentrations at different sizes of squares and different types of
packing on the substrate are presented in table~\ref{tab:pjpc}.
\begin{table}[!htbp]
\caption{Jamming $p_j$ and percolation $p_{\mathrm{c}}$ concentrations for RSA packing of squares
in square lattice (SL) and continuous (C) systems.}\label{tab:pjpc}
\vspace{2ex}
\begin{center}
\begin{minipage}[t]{0.64\textwidth}
\begin{center}
  \begin{tabular}{|c|c|c|c|c|}
\hline
Model                   &System &  $p_j$        &  $p_{\mathrm{c}}$    & Reference \\
\hline\hline
$1 \times 1$, RSA, NN\footnote{Nearest-neighbor exclusion }    &  SL   &  $0.36413$    &  $-$      &~\cite{Dickman1991}   \\
\hline
$1 \times 1$, RSA, NNN\footnote{Nearest- and next-nearest-neighbour exclusion}   &  SL   &  $0.18698$    &  $-$      &~\cite{Dickman1991}   \\
\hline
$1 \times 1$, RSA            &  SL   &  $1$          &$0.592746$ &~\cite{Jan1999}       \\
\hline
$2 \times 2$, RSA            &  SL   &  $0.74788$    &$0.601$    &~\cite{Dickman1991}   \\
\hline
$3 \times 3$, RSA            &  SL   &  $0.681$      &$0.621$    &~\cite{Nakamura1987}  \\
\hline
$4 \times 4$, RSA            &  SL   &  $0.646$      &$-$        &~\cite{Nakamura1987}  \\
\hline
$k \times k$, RSA            &  SL   &  $0.564$      &$-$        &~\cite{Nakamura1987}  \\
\hline
O, RSA\footnote{Oriented}                   &  C    &  $0.562009$   &$-$        &~\cite{Brosilow1991}    \\
\hline
O, RLP\footnote{Random loose packing }                   &  C    &  $0.75$       &$-$        &~\cite{Aristoff2009}    \\
\hline
O, RSA\footnote{Each square has a single chance of adsorption}                   &  C    &  $0.327$      &$-$        &~\cite{Vieira2011}    \\
\hline
RO, RSA\footnote{Randomly oriented}                  &  C    &  $0.532$      &$-$        &~\cite{Vigil1989}     \\
\hline
  \end{tabular}
\end{center}
\end{minipage}
\end{center}
\end{table}

The percolation in the mixtures of squares with different sizes was also intensively studied~\cite{Nakamura1984, Nakamura1985, Sahara1999, Sahara1999a, Lebovka2006, Shida2009a, Shida2010a}. The percolation $k \times k$ squares at $k\geqslant 4$ can be restored by adding  supplementary $(k-1) \times (k-1)$ squares of smaller size to the jammed system. The calculations have shown that percolation threshold is $p_{\mathrm{c}}\approx 0.73$ at large $k$ ($k \geqslant 15$)~\cite{Nakamura1984,Nakamura1985}. The study of percolation in the mixtures of monomers ($k=1$) with $2\times 2$ and $4 \times 4$ squares has shown~\cite{Sahara1999,Sahara1999a} that percolation threshold has increased compared with its value for the ordinary percolation of monomers, $p_{\mathrm{c}}=0.592746$~\cite{Jan1999}. E.g., for the mixture of monomers ($k=1$) and $2\times 2$ squares at equal fractions of the total area, occupied by $1\times1$ and $2\times2$ squares, the percolation concentration was $0.715\pm 0.05$~\cite{Sahara1999,Sahara1999a}.
The model of composite containing monomers (conductors) and $k \times k$ squares (insulators) that fill the space in
a regular manner was shown to be useful in explaining the experimental data on percolation in segregated polymers~\cite{Lebovka2006}.

The present work is devoted to the study of jamming and percolation of the equal size $k \times k$ squares (E-problem) and their mixtures with smaller $m \times m$ squares (M-problem) in a single-cluster growth model.  The Leath--Alexandrowicz (LA) method was used to grow a cluster from an active seed site~\cite{Leath1976, Alexandrowicz1980}. LA method was intensively used to study the ordinary percolation problem for monomers. The deposition rules in a single cluster model are obviously different from those in RSA model and it is expected that percolation and jamming behaviors in single cluster and RSA models should be quite different.  The remainder of the paper is organized as follows. In section~\ref{sec:model}, we describe our model and the details of simulation. The obtained results are discussed in section~\ref{sec:results}. We summarize the results and conclude our paper in section~\ref{sec:concl}.

\section{\label{sec:model}Description of models and details of simulations}

\begin{figure}[!ht] 
\centering
\includegraphics[clip=on, width=0.65\linewidth]{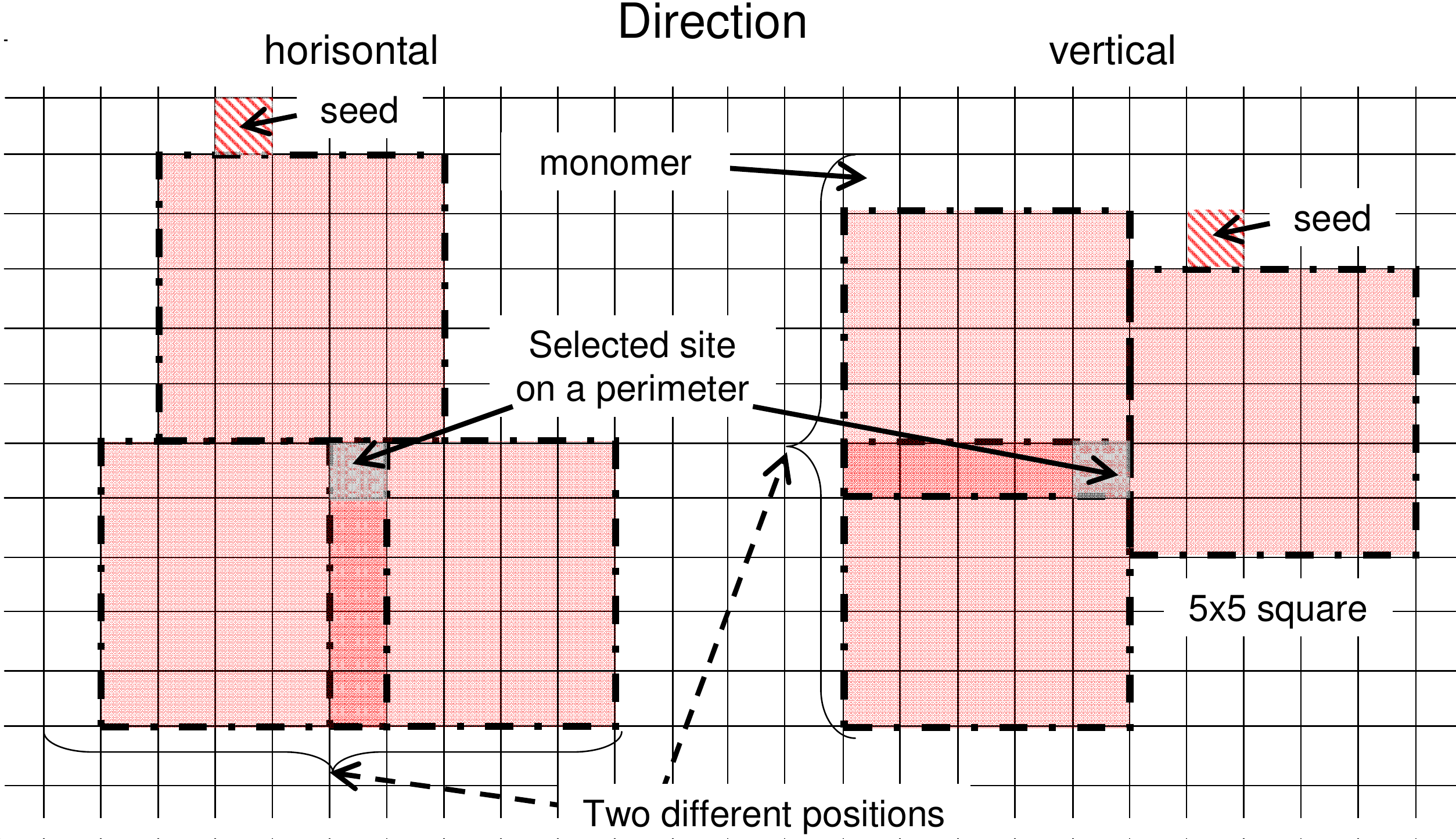}
\caption {\label{fig:LAalgorithm}(Color online) To the description of the computation LA algorithm to grow clusters of $k \times k$ squares (E-problem). }
\end{figure}

The Leath-Alexandrowicz (LA) method~\cite{Leath1976, Alexandrowicz1980} was used to grow a cluster from an active monomer seed on the square lattice. The lattice sites were occupied by addition of the equal size $k \times k$ squares (E-problem) or a mixture of $k \times k$ and $m \times m$ ($m\leqslant  k$) squares (M-problem). In M-problem, $k \times k$ and $m \times m$ squares were assumed to be active (conductive) and blocked (non-conductive), respectively. Note that the simplest case of $m=1$ corresponds to the mixture of $k \times k$ squares and monomer. In order to grow the cluster of equal size $k \times k$ squares (E-problem), LA algorithm uses the following steps (see figure~\ref{fig:LAalgorithm}):
\begin{enumerate}
 \item Occupy an initial seed by a monomer. It has 4 initial perimeter sites;
 \item Deposit randomly the first $k \times k$ square attached to this seed. Determine new perimeter sites;
 \item Randomly choose one perimeter site and try to fill the lattice sites with a new $k \times k$ square. This can be done by two equiprobable ways in horizontal or vertical directions (figure~\ref{fig:LAalgorithm}). In the case of unsuccessful attempt, continue step 3;
 \item Denominate the new added $k \times k$ square as active with probability $\rho$  and as blocked with probability $1-\rho$. Add an active $k \times k$ square to the cluster and determine new perimeter sites. The sites, occupied by the blocked $k \times k$ square, are not tested again;
 \item Continue steps 3 and 4 until there remain no untested perimeter sites.
\end{enumerate}

The time of grown $t$ was evaluated as the number of deposited $k \times k$ squares.

The similar LA algorithm was used for the M-problem [for mixtures of $k \times k$ and $m \times m$ squares ($m\leqslant  k$)].  The relative fraction of active $k \times k$ squares (i.e., the ratio of the area occupied by $k \times k$ squares and the total area occupied by $k \times k$ and $m \times m$ squares) was calculated as follows:
\begin{equation}\label{eq:p}
p=\frac{\rho k^2}{\rho k^2+(1-\rho)m^2}=\left[1-(1/\rho-1)m^2/k^2\right]^{-1}.
\end{equation}

Note, that for the E-problem, $m=k$ and $p=\rho$  is the fraction of active squares.

\begin{figure*}[!ht]
\centering
{\includegraphics*[clip=on,width=0.32\linewidth]{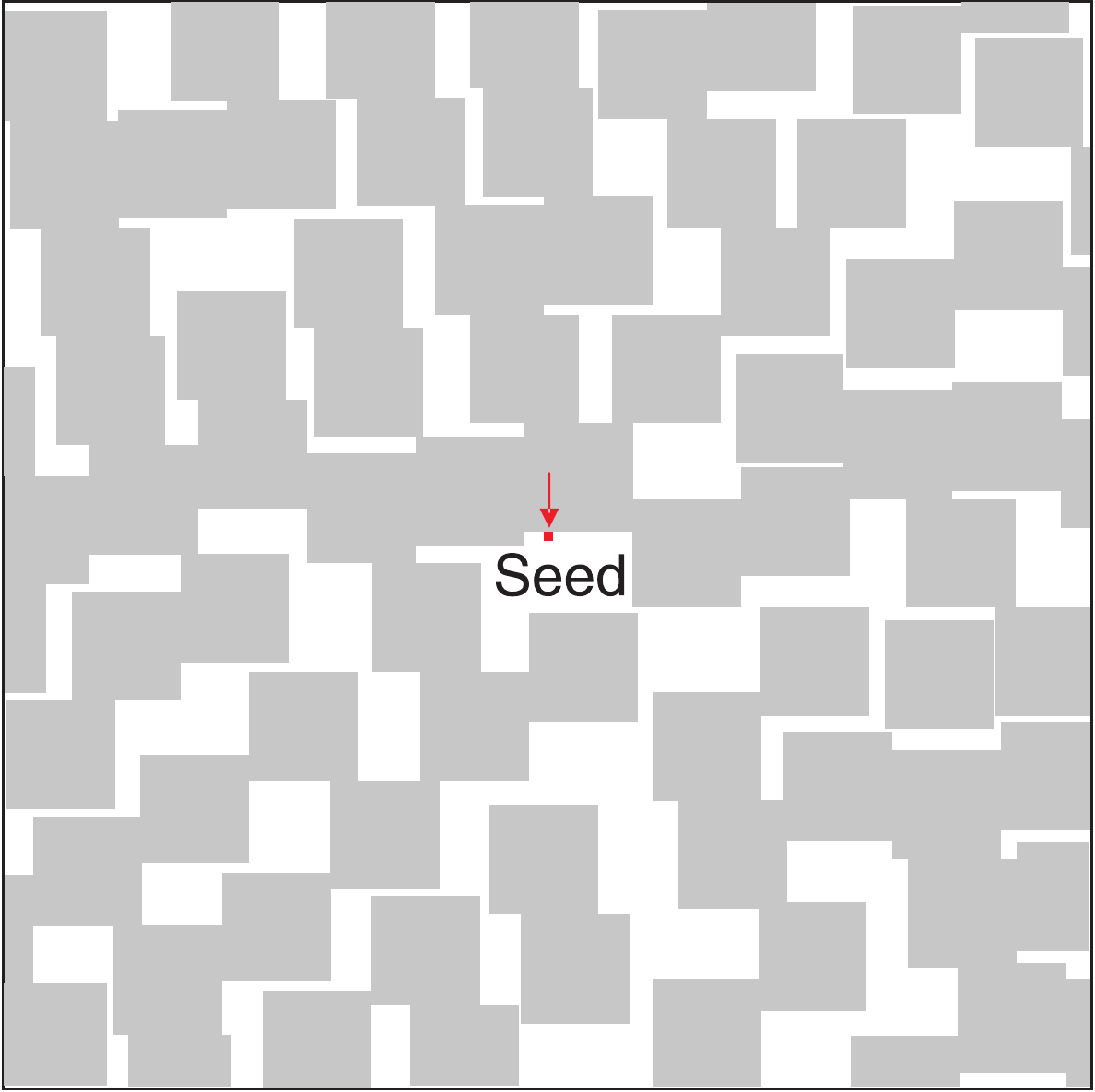}}\hfill
{\includegraphics*[clip=on,width=0.32\linewidth]{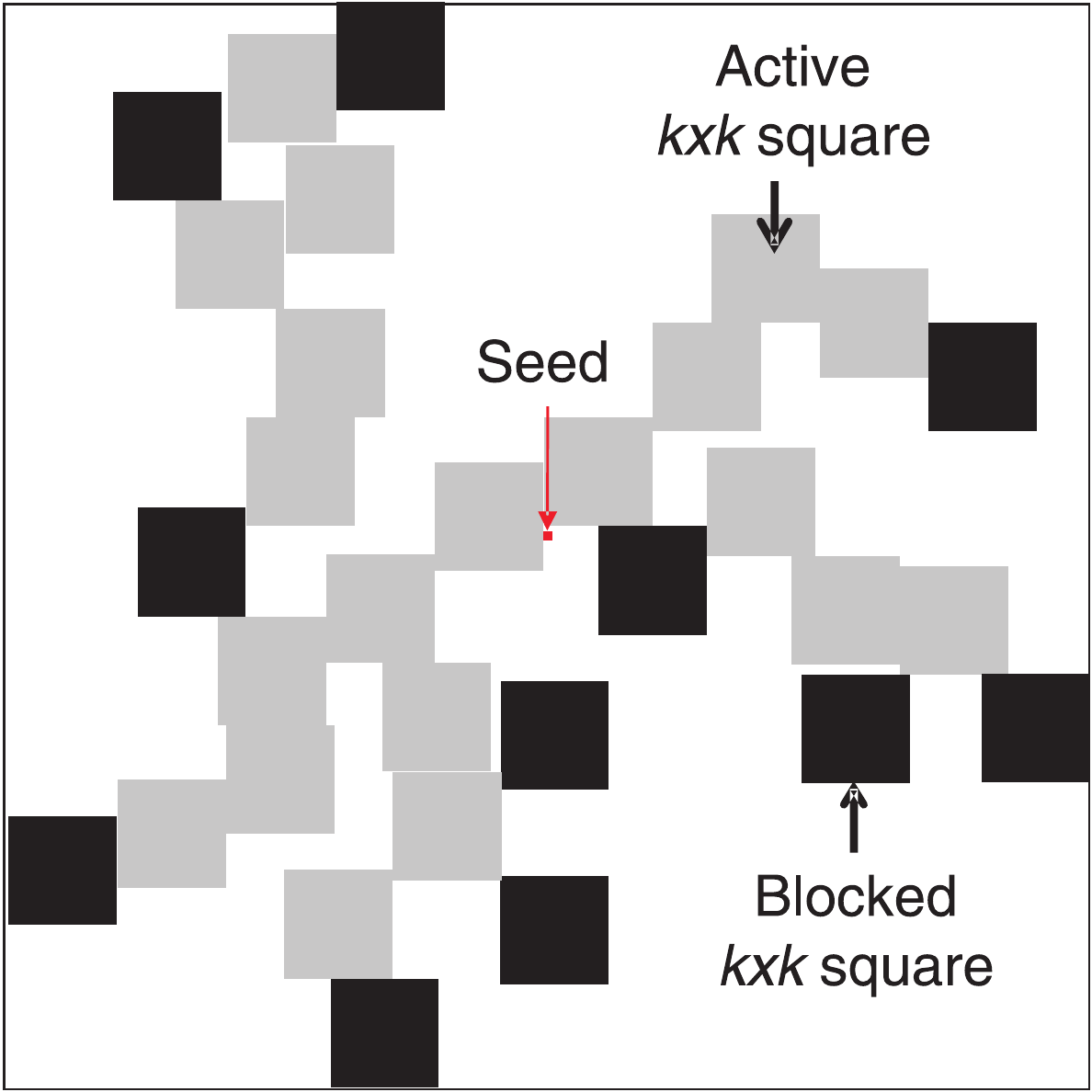}}\hfill
{\includegraphics*[clip=on,width=0.32\linewidth]{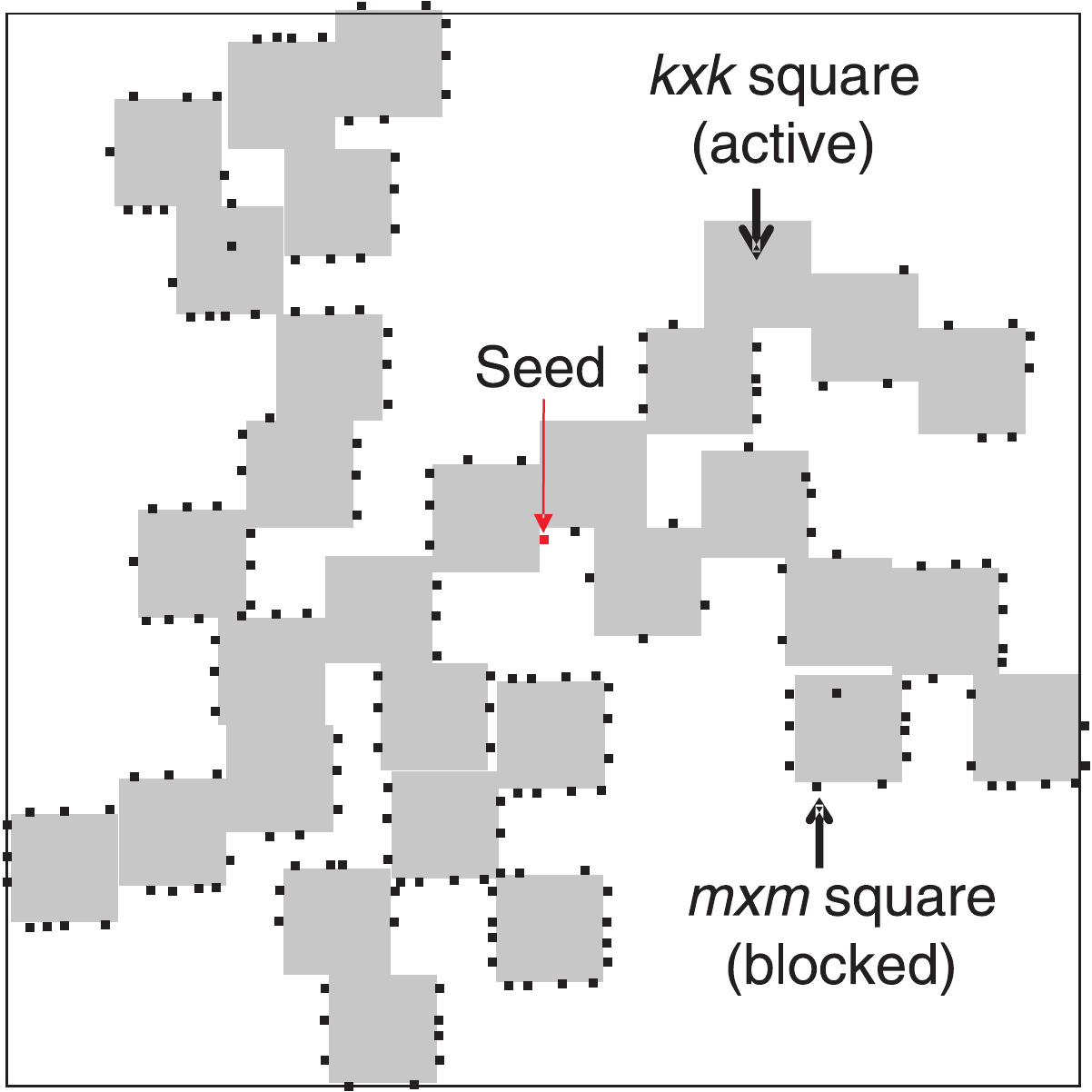}}\\
\caption {\label{fig:clusters}(Color online) Examples of clusters for E-problem [equal size $k \times k$ squares, $p=0$ (a) and $p\neq0$ (b)] and M-problem [a mixture of $k \times k$ and $m \times m$ ($m\leqslant  k$) squares (c)]. Here, gray squares are active (conductive) and black squares are blocked (non-conductive). }
\end{figure*}
Examples of clusters of $k \times k$ squares for E-problem are presented in figure~\ref{fig:clusters}~(a)
($p=0$) and figure~\ref{fig:clusters}~(b) ($p\neq0$). For Monte Carlo simulations with $p=1$, the deposition was terminated when the jamming state was reached, and these data were used to calculate the jamming concentration $p_j$. The percolation concentration $p_{\mathrm{c}}$ was estimated as the threshold concentration that divides the regimes of infinite and finite clusters growth. At $p<p_{\mathrm{c}}$, only finite clusters were grown.

Example of a cluster of $k \times k$ squares, blocked by smaller $m \times m$ squares, is presented in figure~\ref{fig:clusters}~(c).

The E- and M-problems with $k = 2-64$ and $m=1,2,4,8$ on a square lattice of $L \times L$ size have been studied.  The random number generator of Marsaglia et al. was used in these studies~\cite{Marsaglia1990}. The numerical data of Monte Carlo simulations were analyzed for different simulations by the finite-size scaling of a linear lattice with size $L$ varied from $128$ to $8192$. The data were averaged using $1000$ independent runs for $L\leqslant 2048$ and 100--500 runs for larger systems.

\section{\label{sec:results}Results and discussion}

\subsection{\label{subsec:E-problem} Equal sized $k \times k$ squares (E-problem)}
\begin{figure}[!htbp] 
\centering
\includegraphics[clip=on, width=0.65\linewidth]{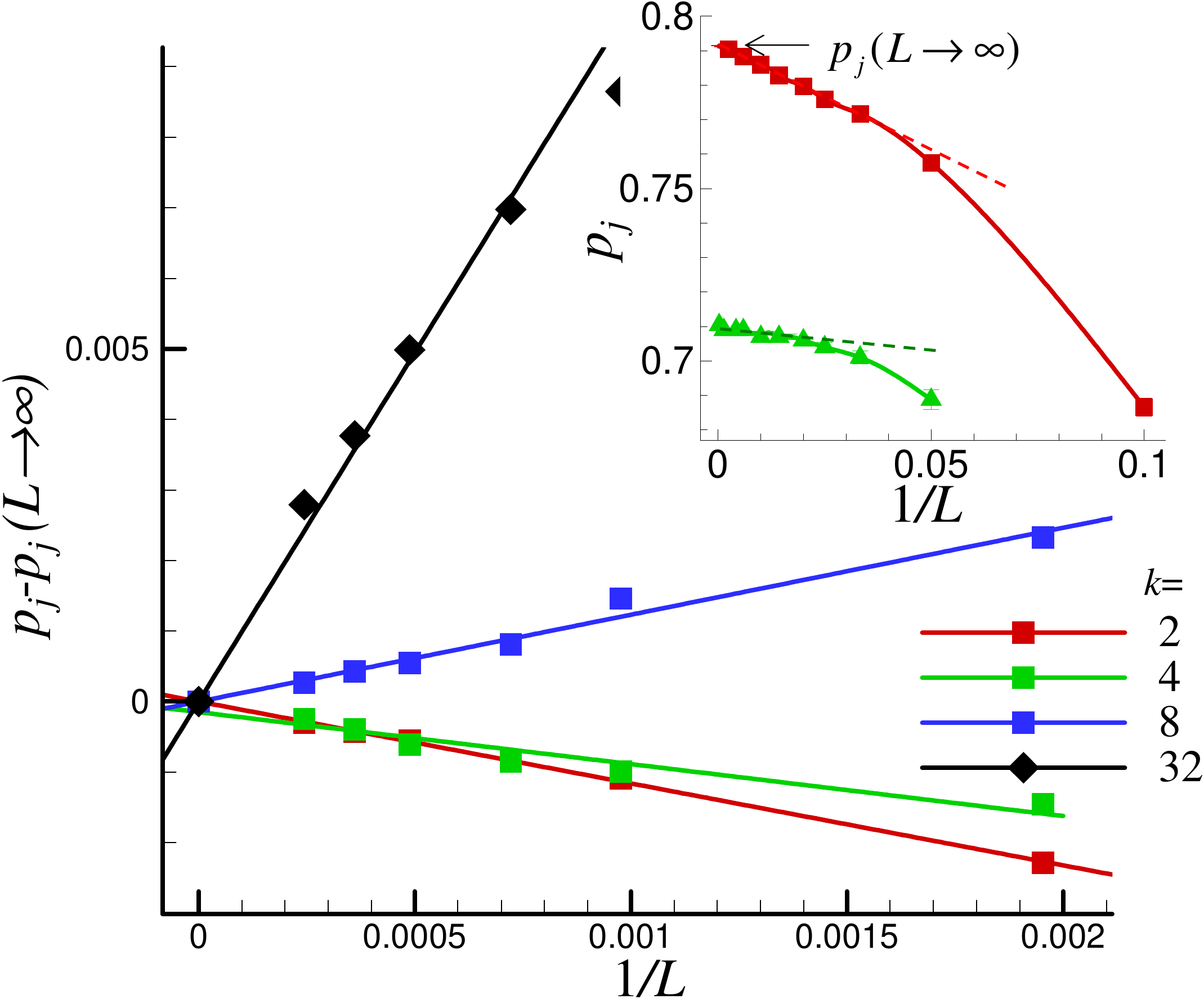}
\caption {\label{fig:FiniteScPj}(Color online) Examples of the finite scaling analysis of jamming concentrations $p_j$ at different sizes of the square $k$. Here,  $p_j(L\rightarrow\infty)$ is the thermodynamic limit ($L\rightarrow \infty$) of the jamming concentration $p_j$ for the single-cluster growth model of equal-sized $k \times k$ squares (E-problem). The inset presents $p_j$ versus $1/L$ dependencies for $1/L\leqslant 0.1$. Error bars are smaller than the symbols.}
\end{figure}

Figure~\ref{fig:FiniteScPj} presents examples of the finite scaling analysis of jamming concentrations $p_j$ at different values of $k$. At large size of the lattice ($L>20 k$), the observed dependencies between $p_j$ and $1/L$ were practically linear (see, inset to figure~\ref{fig:FiniteScPj}) which was similar to the observation for RSA problem of $k \times k$ squares on a square lattice~\cite{Nakamura1986, Nakamura1986a}. To demonstrate the finite-size effects on the jamming concentration more clearly, the results are represented in the form of $p_j(L)-p_j(L\rightarrow\infty)$ versus the inverse lattice size $1/L$.

\begin{figure}[!htbp] 
\centering
\includegraphics[clip=on, width=0.65\linewidth]{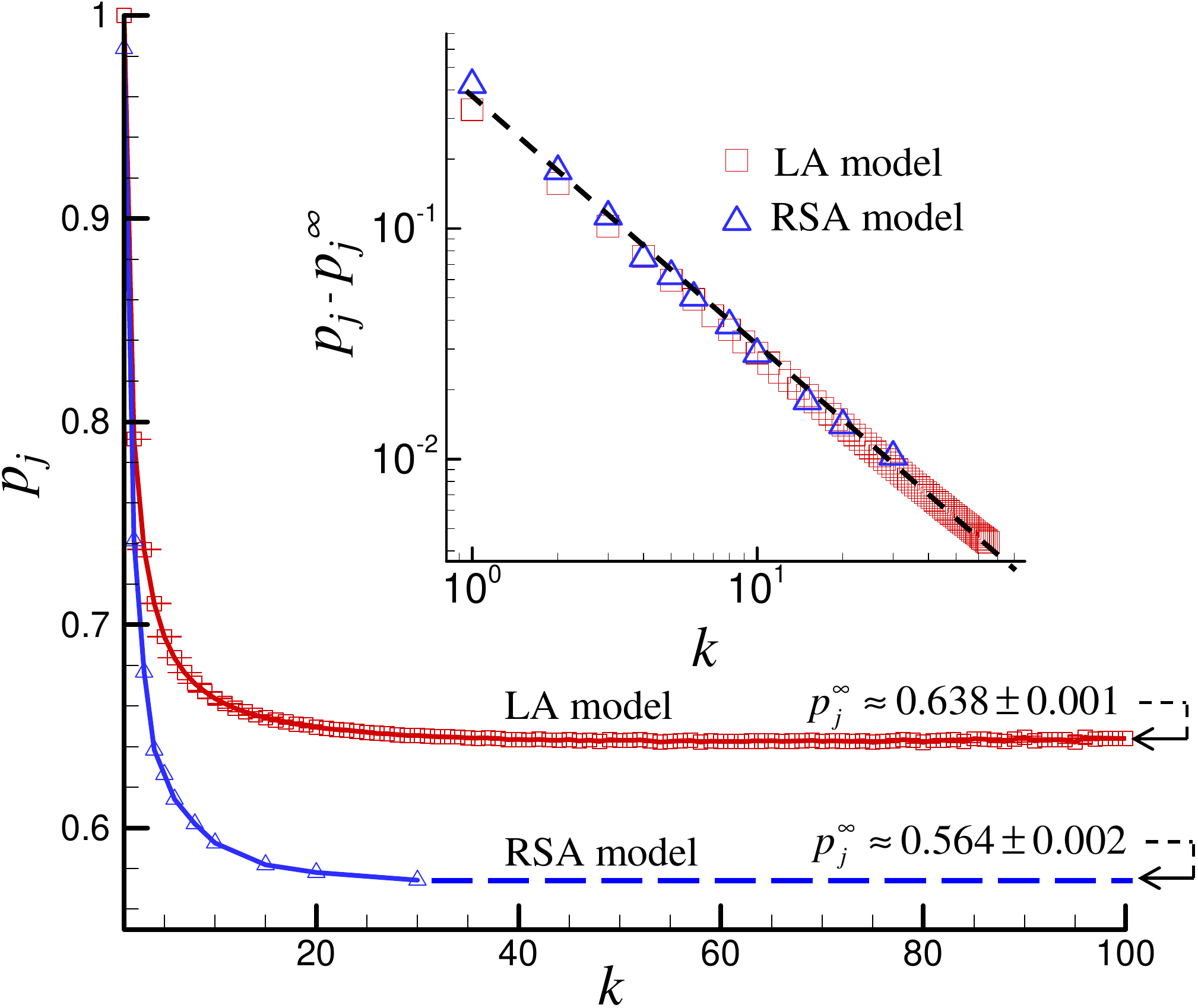}
\caption {\label{fig:Jamming}(Color online) The jamming concentration in the thermodynamic limit ($L\rightarrow\infty$), $p_j$, versus the size of the square, $k$. The data are presented for the single-cluster growth model (E-problem) and RSA model (the data extracted from~\cite{Nakamura1986, Nakamura1986a} of deposition of the equal-sized $k \times k$ squares). The inset presents $p_j -p_j^\infty$  versus $k$ dependencies for single cluster and RSA models. Here, $p_j^\infty$  is the limiting value of $p_j$  at $k\rightarrow\infty$. The dashed line corresponds to the slope of $-1$.  Error bars are smaller than the symbols.}
\end{figure}

Figure~\ref{fig:Jamming} shows the jamming concentration in the thermodynamic limit ($L\rightarrow\infty$), hereinafter referred to as $p_j$, versus the size of the square, $k$. It has been found that numerical results may be rather well fitted by a power law function, equation~(\ref{eq:Nakamura}), with parameters $p_j^\infty = 0.638 ± 0.001$ and   $\alpha=1.053± 0.002$.

\begin{figure}[!htbp] 
\centering
\includegraphics[clip=on, width=0.65\linewidth]{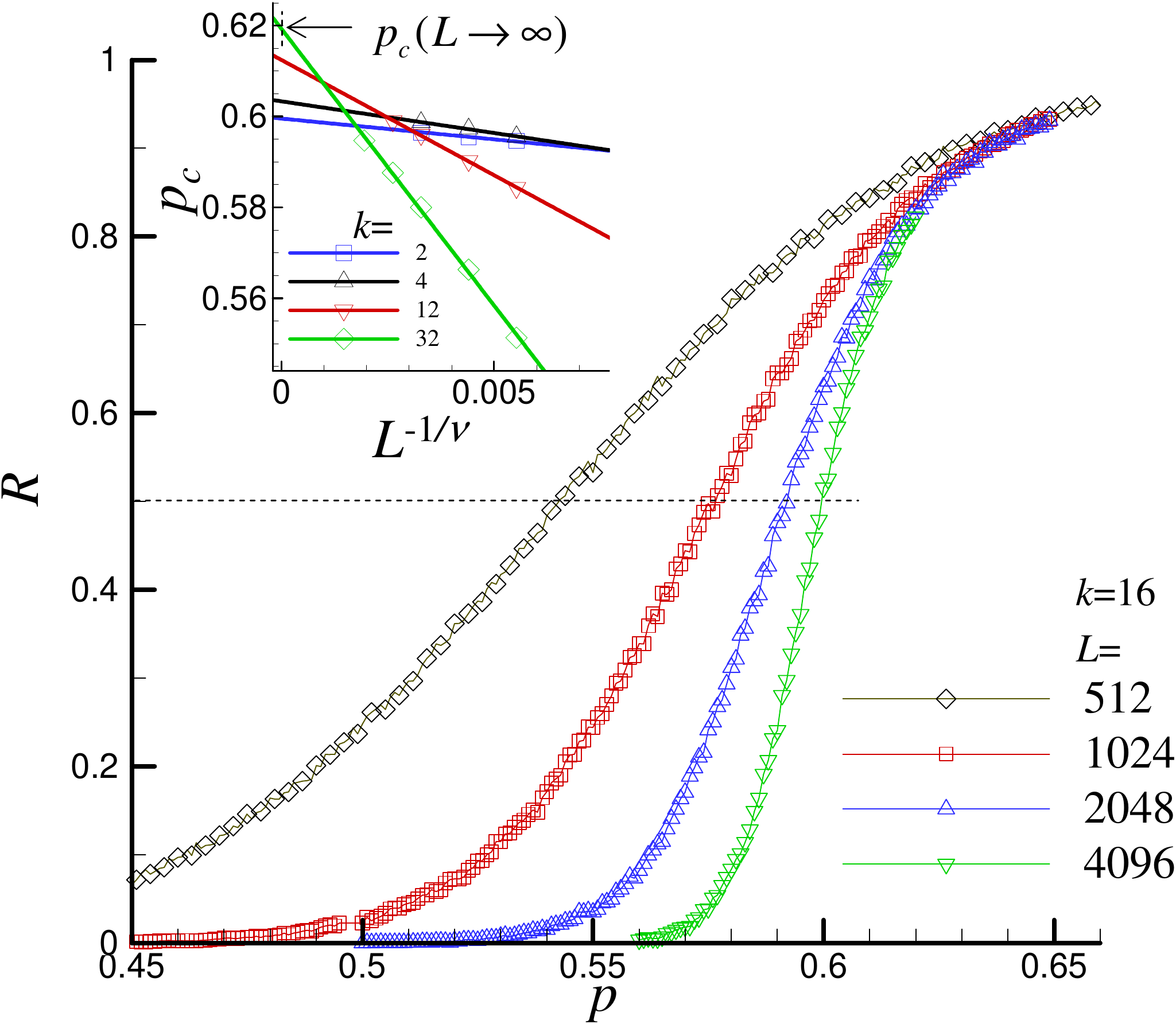}
\caption {\label{fig:RpE}(Color online) Examples of the percolation probability $R$ versus
the fraction of active squares $p$ for the single-cluster growth model of equal size
$k \times k$ squares (E-problem, $k=16$) and different values of $L$. The inset shows
$p_{\mathrm{c}}$ versus $L^{-1/\nu}$ dependencies for different values of $k$.
Here, $\nu = 4/3$ is the critical exponent of correlation length for the ordinary 2D
random percolation problem~\cite{Stauffer1992}. Error bars are smaller than the symbols.}
\end{figure}

The inset to figure~\ref{fig:Jamming} presents $p_j -p_j^\infty$  versus $k$ dependencies for single cluster and RSA models, where the $p_j(k)$ data for RSA model were extracted from~\cite{Nakamura1986,Nakamura1986a}. Here, the dashed line corresponds to the slope of $-1$. This evidences that the jamming concentration $p_j$ is in inverse proportion to the size of the square $k$ for both the single-cluster growth and RSA models.

However, at the same values of $k$, the values of $p_j$ were larger for single-cluster model than for RSA model. This reflects that single-cluster growth rules give more compact packing than RSA deposition rules.

Figure~\ref{fig:RpE} shows the examples of the percolation probability $R$ versus the fraction of active squares $p$ for the single-cluster growth model of equal size $k \times k$ squares ($k=16$) and different values of $L$. To extrapolate estimations of the percolation thresholds $p_{\mathrm{c}}(L)$, obtained at the lattice of size $L$, to the infinitely large lattice $p_{\mathrm{c}}$, the usual finite-size scaling analysis of the percolation behavior was done. To perform extrapolation, the scaling relation
\begin{equation}\label{eq:Pc}
\left|p_{\mathrm{c}} -p_{\mathrm{c}}^\infty\right|\propto k^{-1/\nu},
\end{equation}
was used.

\begin{figure}[!htbp] 
\centering
\includegraphics[clip=on, width=0.65\linewidth]{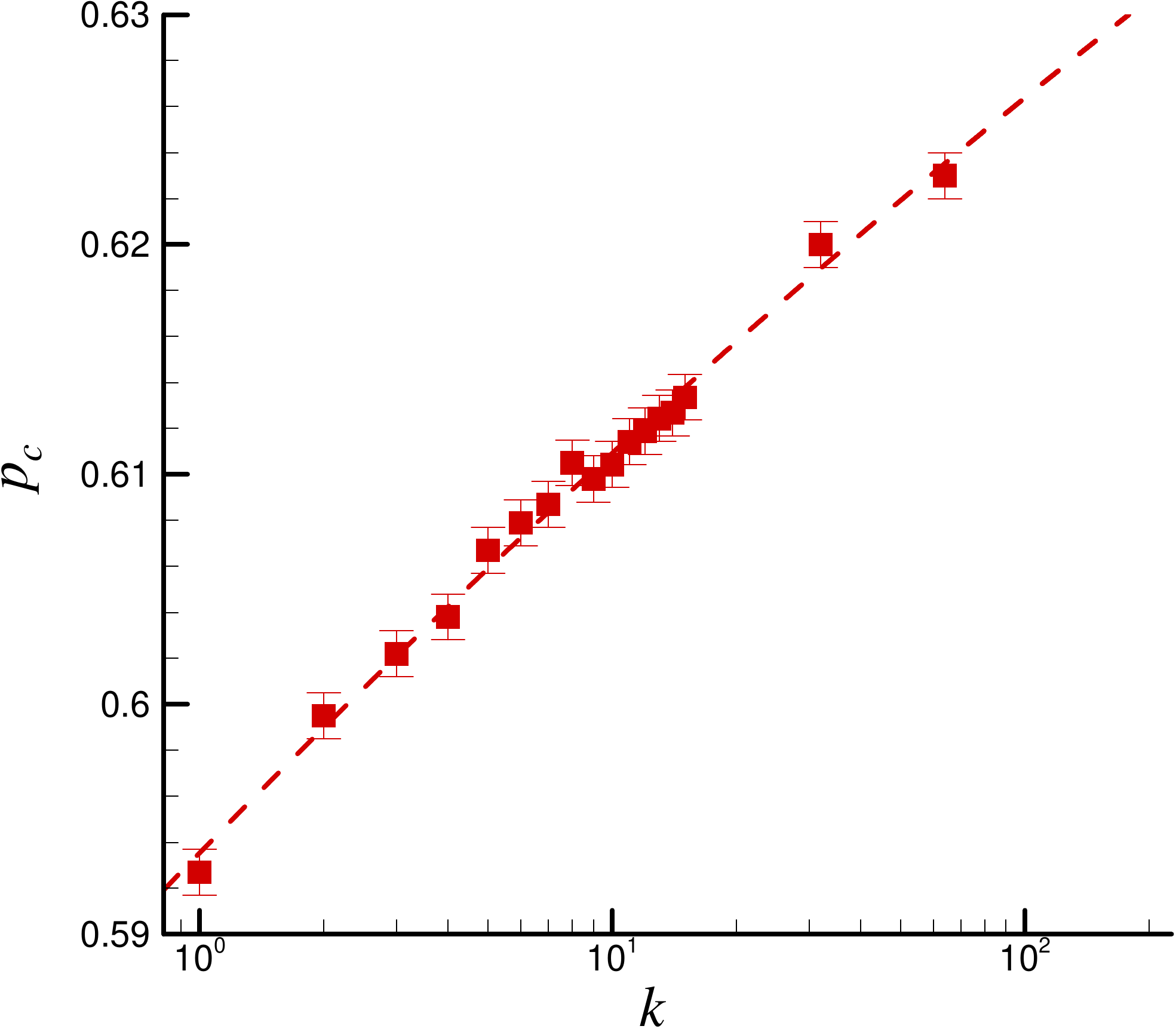}
\caption {\label{fig:FractionPcE}(Color online) Fraction of active squares at the percolation point $p_{\mathrm{c}}$ versus the size of a square $k$ for the single-cluster growth model of equal size $k \times k$ squares (E-problem).}
\end{figure}

Here, $\nu = 4/3$ is the critical exponent of correlation length for the ordinary 2D random percolation problem~\cite{Stauffer1992}. In our study, the typical values of lattice size were $L\leqslant  100k$. The inset to figure~\ref{fig:RpE} shows $p_{\mathrm{c}}$ versus $L^{-1/\nu}$ dependencies for different values of $k$. Our results evidence that the problem studied belongs to universality of the ordinary 2D random percolation problem at different values of $k$~\cite{Stauffer1992}.

Figure~\ref{fig:FractionPcE} presents the percolation threshold $p_{\mathrm{c}}$ versus the size of a square $k$ for the single-cluster growth model of equal size $k \times k$ squares. The value of $p_{\mathrm{c}}$ continually increased with an increase of $k$ in the studied range of $k=2\dots 64$. In the problem under consideration, the formation of the percolation cluster reflects the connectivity between the central seed and infinitely distant $k \times k$ squares through the network of active $k \times k$ squares, and each blocked $k \times k$ square terminates the growth of the cluster in the vicinity of this square. It may be assumed that this connectivity is similar to that in the Bethe lattice in the limit of $k\rightarrow\infty$. The percolation threshold for the Bethe lattice can be calculated as $p_{\mathrm{c}}=1/(z-1)$ where $z$ is the coordination number. In the limit of $k\rightarrow\infty$, the studied E-problem corresponds to a continuous (off-lattice) problem with the maximum coordination number $z=4$. Thus, it may be speculated that for our E-problem, $p_{\mathrm{c}}^\infty\approx 0.75$.

Dashed line in figure~\ref{fig:FractionPcE} corresponds to the following power relation:
\begin{equation}\label{eq:Pck}
p_{\mathrm{c}}=0.75 -a k^{-\alpha},
\end{equation}
where $a=0.156\pm 0.001$ and $\alpha=0.051\pm 0.001$.

We see that computational data points can be satisfactorily fitted using equation~(\ref{eq:Pck}). However, larger scale computations are required in future in order to make a more precise estimation of the value.

\subsection{\label{subsec:M-problem} Mixture of $k \times k$ and $m \times m$ ($m\leqslant  k$) squares (M-problem)}

\begin{figure}[!ht] 
\centering
\includegraphics[clip=on, width=0.65\linewidth]{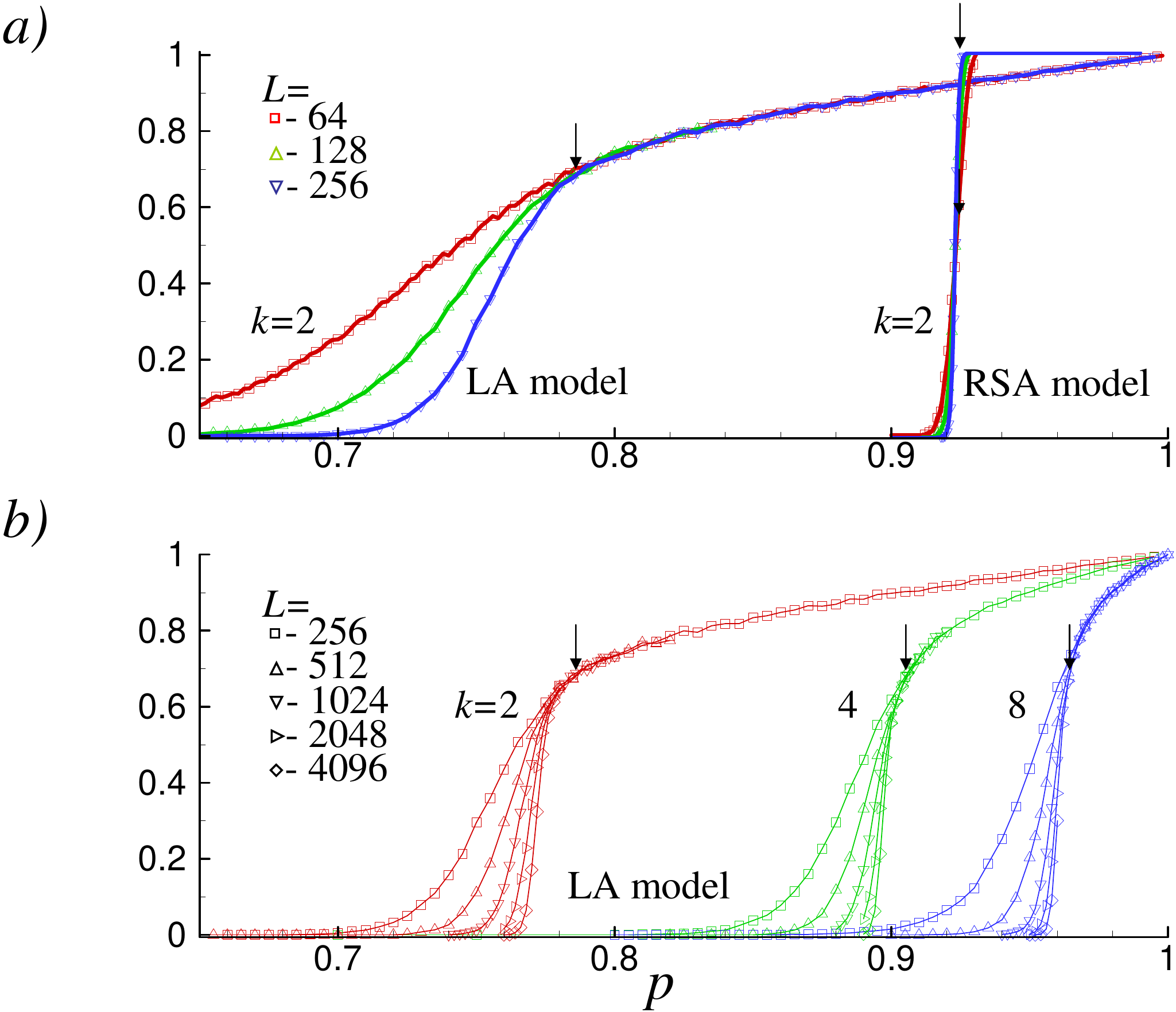}
\caption {\label{fig:RpM}(Color online) Examples of the percolation probability $R$ versus the fraction of active $k \times k$ squares $p$ for mixture of $k \times k$ squares and monomers. Here, the $R(p)$ dependencies are compared for the single-cluster LA growth model (M-problem) and RSA model ($\rho=0.95$) (a) for $k=2$ (a) and for LA growth model (M-problem), $k=2, 4, 8$ (b). The data are presented for different values of $L$.  Error bars are smaller than the symbols.The positions of the percolation threshold are shown by arrows.}
\end{figure}
Figure~\ref{fig:RpM} presents examples of the percolation probability $R$ versus the fraction of active $k \times k$ squares $p$ for the mixture of $k \times k$ squares and monomers. Figure~\ref{fig:RpM}~(a) compares $R(p)$ dependencies for the single-cluster LA growth model (M-problem) and RSA model ($\rho=0.95$) (a) for $k=2$. In RSA model, the inactive (insulating) monomers were initially deposited on the square lattice with probability of $1-\rho$. Then, $k \times k$ squares were deposited using RSA algorithm and the relative fraction of active $k \times k$ squares was calculated using equation~(\ref{eq:p}). It is interesting that the introduction of blocking inactive (insulating) monomers resulted in a noticeable increase of threshold concentration $p_{\mathrm{c}}$. E.g., for $k=2$ in absence of blocking monomers, the value of $p_{\mathrm{c}}$ was $\approx 0.600-0.601$ both for single cluster LA and RSA models. However, in presence of blocking monomers, the value of $p_{\mathrm{c}}$  increased up to $0.782$ and $0.993$, for LA and RSA models, respectively.

\begin{figure}[!htbp] 
\centering
\includegraphics[clip=on, width=0.5\linewidth]{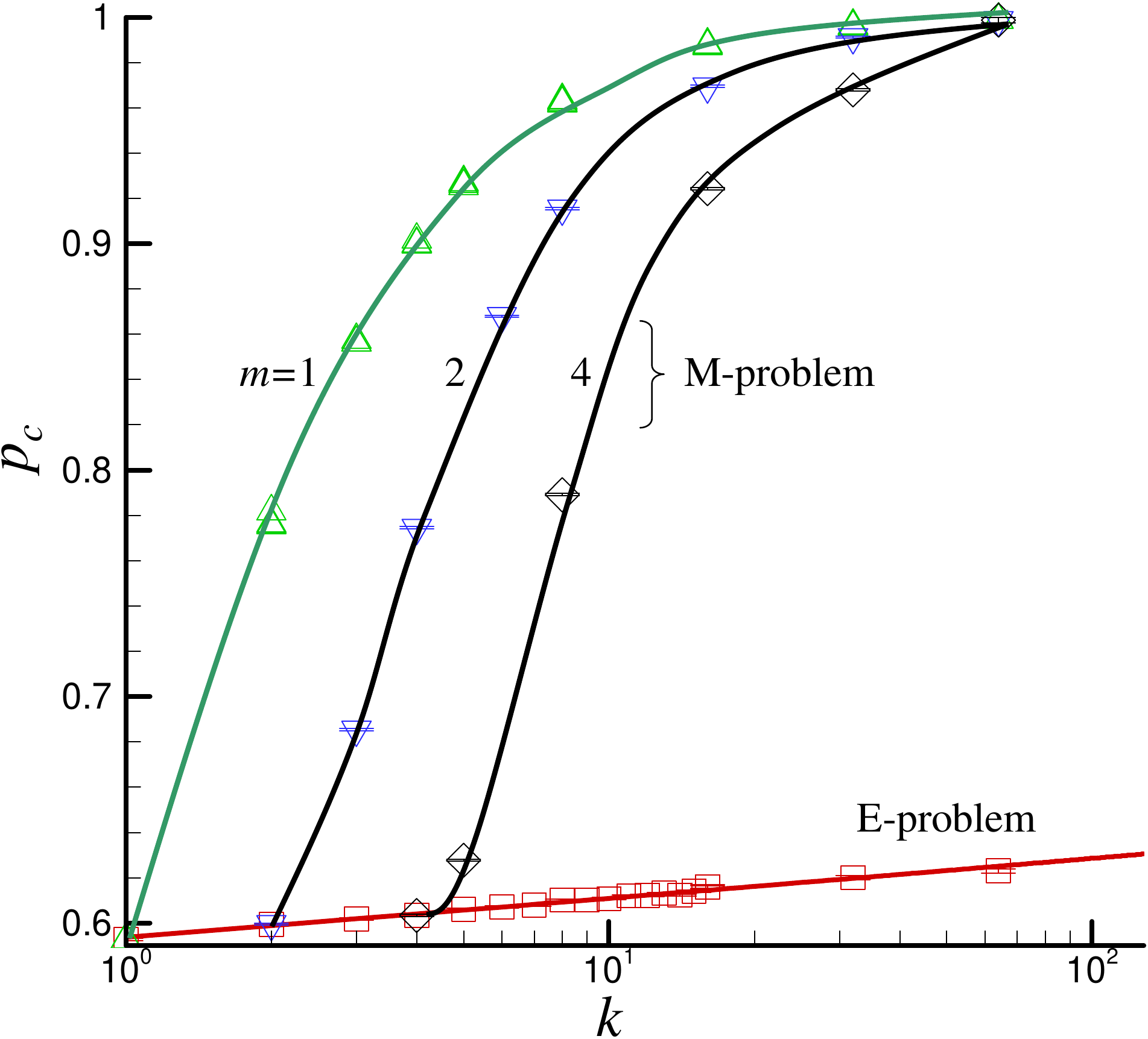}
\caption {\label{fig:FractionPcM}(Color online) Fraction of active $k \times k$ squares at the percolation point $p_{\mathrm{c}}$ versus the size of a square $k$ for E-problem (equal size $k \times k$ squares) and M-problem (a mixture of $k \times k$ and $m \times m$ ($m\leqslant  k$) squares). Error bars are smaller than the symbols. }
\end{figure}

Figure~\ref{fig:RpM}~(b) compares $R(p)$ dependencies for M-problem for different values of $k$.
These data were used for a finite scaling analysis and determination of the percolation threshold, $p_{\mathrm{c}}$ (figure~\ref{fig:FractionPcM}).

The results show that for the mixture of $k \times k$ and $m \times m$ squares, the percolation threshold $p_{\mathrm{c}}$ was increasing noticeably with an increase of $k$ at a fixed value of $m$ (figure~\ref{fig:FractionPcM}). Moreover, it approached  1 at $k \geqslant  10m$. Thus, for the model of the mixture of $k \times k$ and $m \times m$ squares, the percolation of larger active $k \times k$ squares may be suppressed in the presence of considerably smaller blocked $m \times m$ squares. This is a natural account for the effective blocking of the side faces of $k \times k$ squares at $k\gg m$.

\begin{figure}[!htbp] 
\centering
\includegraphics[clip=on, width=0.65\linewidth]{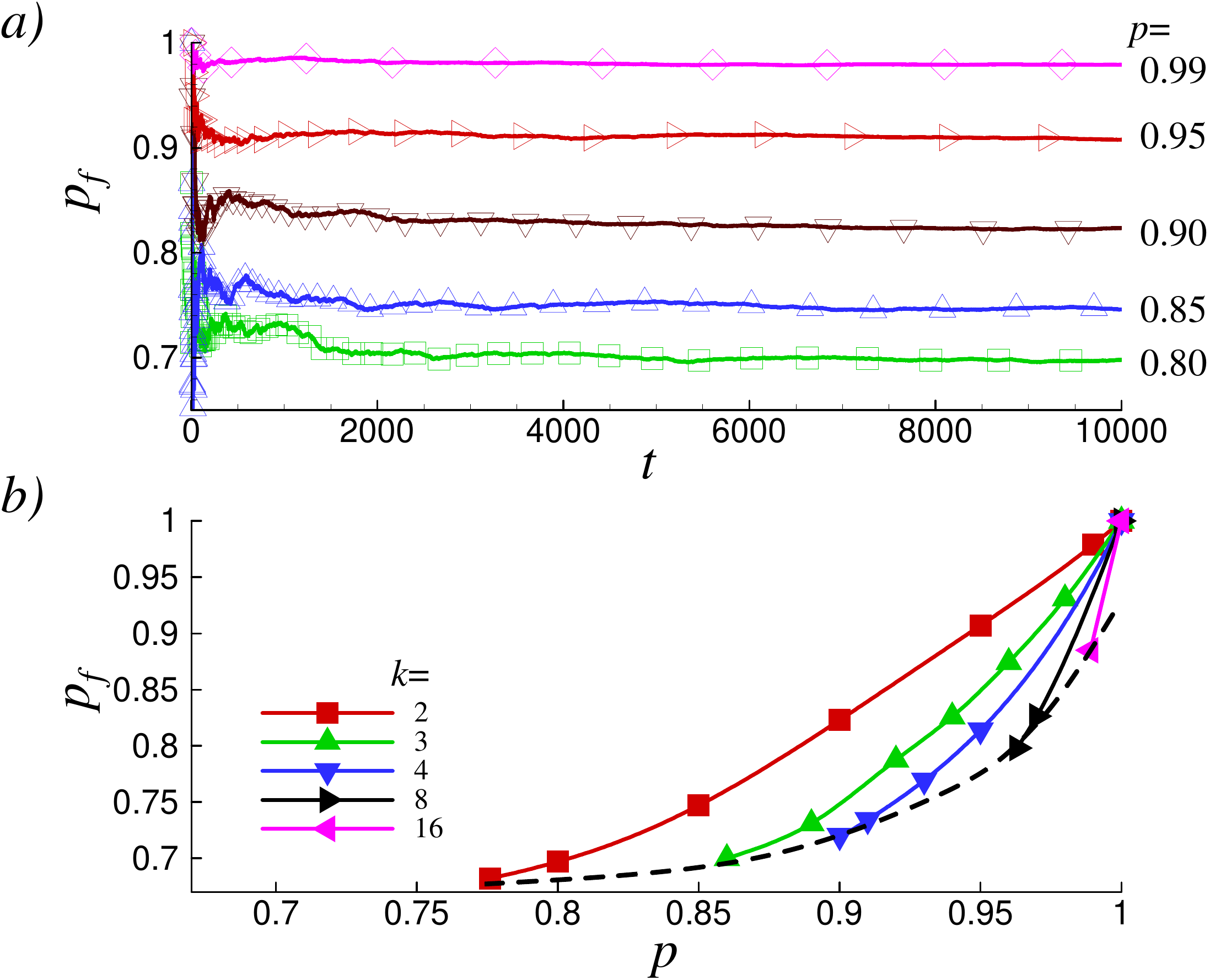}
\caption {\label{fig:PfM}(Color online) Actual value of the fraction of active squares $p_{\mathrm{f}}$ versus the time of growth $t$ at different predefined values of $p$ (a) and values of $p_{\mathrm{f}}$ versus $p$ at different values of $k$. Data are for a mixture of $k \times k$ squares and monomers above the percolation threshold of active squares, i.e., at $p\geqslant  p_{\mathrm{c}}$.}
\end{figure}

Finally, note that detailed experiments revealed the difference between the actual, $p_{\mathrm{f}}$, and predefined, $p$, fractions of active squares for M-problem (a mixture of $k \times k$ and $m \times m$ squares). The Monte Carlo simulation evidences that the actual value $p_{\mathrm{f}}$ was noticeably smaller than the pre-determined $p$ value (figure~\ref{fig:PfM}). This fact may be explained as follows. In the course of the cluster growth, the network of finite size pores was formed. We can expect that the probability of deposition is smaller for $k \times k$ squares due to stronger spatial restrictions and possibility of rejection of the deposition attempts. This results in the reduction of the actual value of $p_{\mathrm{f}}$ in the course of the growth [figure~\ref{fig:PfM}~(a)]. This effect is quite similar to the observed differences between pre-determined and actually observed orientation order parameters in RSA model for partially oriented $k$-mers on a square lattice~\cite{Lebovka2011, Tarasevich2012}. For example, figure~\ref{fig:PfM}~(b) demonstrates that this difference between  $p_{\mathrm{f}}$ and $p$ may be rather noticeable for a mixture of $k \times k$ squares and monomers above the percolation threshold of active squares and it was growing with an increase of the square size $k$.

\section{\label{sec:concl}Conclusion}

In this paper, the jamming and percolation of parallel squares in a single-cluster growth model were investigated by computer simulations. The sites of a square lattice were occupied by addition of equal size $k \times k$ squares (E-problem) or a mixture of $k \times k$ and $m \times m$ ($m\leqslant  k$) squares (M-problem). The larger $k \times k$ squares were assumed to be active (conductive) and the smaller $m \times m$ squares were assumed to be blocked (non-conductive).  For jamming concentration of equal size squares (E-problem), the power low dependence of type $\mid p_j -p_j^\infty \mid\propto k^{-\alpha}$ was obtained, where $p_j^\infty = 0.638 ± 0.001$ and   $\alpha\approx 1.0$. The data also evidence that the studied problem belongs to the universality of ordinary 2D random percolation at different values of $k$. The percolation threshold $p_{\mathrm{c}}$ increased with an increase of $k$. It was speculated that $p_{\mathrm{c}}(k)$ can be described by the relation $p_{\mathrm{c}}=0.75-a k^{-\alpha}$, where $a=0.156 \pm 0.001$ and  $\alpha=0.051 \pm 0.001$. For mixture of $k \times k$ and $m \times m$ ($m\leqslant  k$) squares (M-problem), the percolation threshold $p_{\mathrm{c}}$ increased noticeably with an increase of $k$ at fixed value of $m$ and approached  1 at $k\geqslant  10m$. It was demonstrated that percolation of larger active $k \times k$ squares can be effectively suppressed in  the presence of smaller blocked $m \times m$ squares for the M-problem.

\section{\label{sec:Ack}Acknowledgements}
Authors would like to acknowledge the partial financial support of project 43--02--14(U), Ukraine (N.L.) and of project RFBR 14--02--90402\_Ukr, Russia (Yu.T.). Authors also thank Dr. N.S.~Pivovarova for her help with the preparation of the manuscript.


\newpage
\begin{thebibliography}{10}
\providecommand{\url}[1]{\texttt{#1}}
\providecommand{\urlprefix}{URL }
\providecommand{\eprint}[2][]{\url{#2}}

\bibitem{Wang1998}
Wang J.S., Physica A, 1998, \textbf{254}, No.~1--2, 179; \doi{10.1016/S0378-4371(98)00028-4}.

\bibitem{Privman2000}
Privman V., J. Adhesion, 2000, \textbf{74}, No.~1--4, 421; \doi{10.1080/00218460008034540}.

\bibitem{Budinski2012}
Budinski-Petkovi\'{c} L., Loncarevi\'{c} I., Petkovi\'{c} M., Jaksi\'{c} Z., Vrhovac S., Phys. Rev. E, 2012,
\textbf{85}, No.~6, 061117; \doi{10.1103/PhysRevE.85.061117}.

\bibitem{Evans1993}
Evans J.W., Rev. Mod. Phys., 1993, \textbf{65}, 1281; \doi{10.1103/RevModPhys.65.1281}.

\bibitem{Manna1991}
Manna S., Svraki\'{c} N., J. Phys. A-Math. Gen., 1991, \textbf{24}, No.~12, L671; \doi{10.1088/0305-4470/24/12/003}.

\bibitem{Becklehimer1992}
Becklehimer J., Pandey R., Physica A, 1992, \textbf{187}, No.~1--2, 71; \doi{10.1016/0378-4371(92)90409-J}.

\bibitem{Leroyer1994}
Leroyer Y., Pommiers E., Phys. Rev. B, 1994, \textbf{50}, No.~5, 2795; \doi{10.1103/PhysRevB.50.2795}.

\bibitem{Vandewalle2000}
Vandewalle N., Galam S., Kramer M., Eur. Phys. J. B, 2000, \textbf{14}, No.~3, 407; \doi{10.1007/s100510051047}.

\bibitem{Kondrat2001}
Kondrat G., P\c{e}kalski A., Phys. Rev. E, 2001, \textbf{63}, No.~5, 051108; \doi{10.1103/PhysRevE.63.051108}.

\bibitem{Lebovka2011}
Lebovka N., Karmazina N., Tarasevich Y., Laptev V., Phys. Rev. E, 2011, \textbf{84}, No.~6,
  029902; \\ \doi{10.1103/PhysRevE.84.061603}.

\bibitem{Tarasevich2012}
Tarasevich Y., Lebovka N., Laptev V., Phys. Rev. E, 2012, \textbf{86}, No.~6, 061116; \doi{10.1103/PhysRevE.86.061116}.

\bibitem{Longone2012}
Longone P., Centres P., Ramirez-Pastor A., Phys. Rev. E, 2012, \textbf{85}, No.~1, 011108; \doi{10.1103/PhysRevE.85.011108}.

\bibitem{Pawlowska2012}
Paw{\l}owska M., {Z}erko S., Sikorski A., J. Chem. Phys., 2012,
  \textbf{136}, No.~4, 2012; \doi{10.1063/1.3679168}.


\bibitem{Pawlowska2013}
Paw{\l}owska M., Sikorski A., J. Mol. Model., 2013,
  \textbf{19}, No.~10, 4251; \doi{10.1007/s00894-013-1892-y}.

\bibitem{Adamczyk2008}
Adamczyk P., Romiszowski P., Sikorski A., J. Chem. Phys., 2008,
  \textbf{128}, No.~15, 154911; \doi{10.1063/1.2907715}.

\bibitem{Nakamura1987}
Nakamura M., Phys. Rev. A, 1987, \textbf{36}, No.~5, 2384; \doi{10.1103/PhysRevA.36.2384}.

\bibitem{Vigil1989}
Vigil R.D., Ziff R.M., J. Chem. Phys., 1989, \textbf{91},
  No.~4, 2599; \doi{10.1063/1.457021}.

\bibitem{Dickman1991}
Dickman R., Wang J.S., Jensen I., J. Chem. Phys., 1991,
  \textbf{94}, No.~12, 8252; \doi{10.1063/1.460109}.

\bibitem{Connelly2014}
Connelly R., Dickinson W., Philos. Trans. Roy. Soc. A, 2014, \textbf{372}, No.~2008,
  20120039; \doi{10.1098/rsta.2012.0039}.

\bibitem{Zwanzig1956}
Zwanzig R., J. Chem. Phys., 1956, \textbf{24}, 855; \doi{10.1063/1.1742621}.

\bibitem{Hoover1964}
Hoover W.G., J. Chem. Phys., 1964, \textbf{4}, No.~4, 937; \doi{10.1063/1.1725285}.

\bibitem{Carlier1972}
Carlier C., Frisch H.L., Phys. Rev. A, 1972, \textbf{6}, 1153; \doi{10.1103/PhysRevA.6.1153}.

\bibitem{Hoover2009}
Hoover W., Hoover C.G., Bannerman M.N., J. Stat. Phys., 2009,
  \textbf{136}, 715; \doi{10.1007/s10955-009-9795-0}.

\bibitem{Wojciechowski2004}
Wojciechowski K.W., Frenkel D., Comp. Methods Sci. Tech., 2004, \textbf{10}, No.~2, 235; \\ \doi{10.12921/cmst.2004.10.02.235-255}.

\bibitem{Rodgers1993}
Rodgers G.J., Phys. Rev. E, 1993, \textbf{48}, No.~6, 4271; \doi{10.1103/PhysRevE.48.4271}.

\bibitem{Vieira2011}
Vieira M.C., Gomes M., de~Lima J., Physica A, 2011, \textbf{390}, 3404; \doi{10.1016/j.physa.2011.05.025}.

\bibitem{Nakamura1986}
Nakamura M., Phys. Rev. A, 1986, \textbf{34}, No.~4, 3356; \doi{10.1103/PhysRevA.34.3356}.

\bibitem{Nakamura1986a}
Nakamura M., J. Phys. A, 1986,
  \textbf{19}, No.~12, 2345; \doi{10.1088/0305-4470/19/12/020}.

\bibitem{Jan1999}
Jan N., Physica A, 1999,
  \textbf{266}, No.~1--4, 72; \doi{10.1016/S0378-4371(98)00577-9}.

\bibitem{Brosilow1991}
Brosilow B.J., Ziff R.M., Vigil R.D., Phys. Rev. A, 1991, \textbf{43}, 631; \doi{10.1103/PhysRevA.43.631}.

\bibitem{Aristoff2009}
Aristoff D., Radin C., J. Stat. Phys., 2009, \textbf{135}, 1; \doi{10.1007/s10955-009-9722-4}.

\bibitem{Nakamura1984}
Nakamura M., J. Appl. Phys., 1984, \textbf{56}, No.~3, 806; \doi{10.1063/1.334011}.

\bibitem{Nakamura1985}
Nakamura M., J. Appl. Phys., 1985, \textbf{58}, No.~9, 3499; \doi{10.1063/1.335774}.

\bibitem{Sahara1999}
Sahara R., Mizuseki H., Ohno K., Kawazoe Y., Mater. Trans., 1999,
  \textbf{40}, No.~11,  1314; \\\doi{10.2320/matertrans1989.40.1314}.

\bibitem{Sahara1999a}
Sahara R., Mizuseki H., Ohno K., Kawazoe Y., J. Phys. Soc. Jpn., 1999, \textbf{68}, No.~12, 3755; \doi{10.1143/JPSJ.68.3755}.

\bibitem{Lebovka2006}
Lebovka N., Lisunova M., Mamunya Y.P., Vygornitskii N., J. Phys. D, 2006, \textbf{39}, 2264; \\ \doi{10.1088/0022-3727/39/10/040}.

\bibitem{Shida2009a}
Shida K., Sahara R., Tripathi M., Mizuseki H., Kawazoe Y., Mater. Trans., 2009, \textbf{50}, No.~12, 2848; \\\doi{10.2320/matertrans.M2009202}.

\bibitem{Shida2010a}
Shida K., Sahara R., Tripathi M., Mizuseki H., Kawazoe Y., Mater. Trans., 2010, \textbf{51}, No.~6, 1141; \\ \doi{10.2320/matertrans.M2010019}.

\bibitem{Leath1976}
Leath P., Phys. Rev. B, 1976, \textbf{14}, No.~11, 5046; \doi{10.1103/PhysRevB.14.5046}.

\bibitem{Alexandrowicz1980}
Alexandrowicz Z., Phys. Lett. A, 1980, \textbf{80}, No.~4, 284; \doi{10.1016/0375-9601(80)90023-7}.

\bibitem{Marsaglia1990}
Marsaglia G., Zaman A., Tsang W., Stat. Probabil. Lett., 1990,
  \textbf{9}, No.~1, 35; \doi{10.1016/0167-7152(90)90092-L}.

\bibitem{Stauffer1992}
Stauffer D., Aharony A., Introduction to Percolation Theory, Taylor \& Francis,
  London, 1992.

\end{thebibliography}

\newpage

\ukrainianpart
\title{Джамінг та перколяція паралельних квадратів в однокластерній моделі росту}
\author{І.О. Крючевський\refaddr{label1}, Л.А. Булавін\refaddr{label1}, Ю.Ю. Тарасевич \refaddr{label2}, М.І. Лебовка\refaddr{label3} }
\addresses{
\addr{label1} Київський національний універститет ім. Тараса Шевченка, фізичний факультет, \\
пр. академіка Глушкова,  2, 03127 Київ,  Україна,
\addr{label2} Астраханський державний університет,  вул. Татіщева, 20a, 414056 Астрахань,  Росія
\addr{label3} Інститут біоколоїдної хімії ім. Ф.Д. Овчаренка НАН України, \\ бульв. академіка Вернадського, 42, 03142 Київ,  Україна
}
%
%
%

\makeukrtitle

\begin{abstract}
\tolerance=3000%
В роботі вивчено явища джамінгу і перколяції паралельних квадратів для однокластерної моделі росту.
Для росту кластеру з активного зародку використовувався метод Ліса-Александровича.
Вузли квадратної ґратки займалися додаванням однакових $k \times k$ квадратів (E-задача)
або суміші  $k \times k$ і $m \times m$ ($m \leqslant  k$) квадратів (M-задача).
Припускалося, що більші $k \times k$ області були активними (провідними), а менші
були заблокованими (непровідними).  Для $k \times k$ квадратів однакового розміру (E-задача)
за умови $k\rightarrow\infty$ було отримано таке значення концентрації джамінгу $p_j = 0.638 ± 0.001$ .
Це значення було істотно меншим за  отримане раніше для  моделі випадкової послідовної адсорбції:  $p_j = 0.564 ± 0.002$.
Було показано, що величина перколяційного порогу $p_{\mathrm{c}}$ (тобто відношення
площі активних $k \times k$ квадратів до загальної площі осаджених $k \times k$
квадратів в перколяційній точці) зростала при збільшенні $k$. Для суміші $k \times k$ і $m \times m$
квадратів (M-задача) величина $p_{\mathrm{c}}$ сильно зростала при збільшенні $k$ при фіксованому
значенні $m$ та наближалась до 1 при$k\geqslant  10m$. Це пов'язано з тим, що перколяція більших
активних квадратів для M-задачі може ефективно пригнічуватися за наявності невеликої кількості малих заблокованих квадратів.
\keywords джамінг, перколяція, квадрати, невпорядковані системи, метод Монте Карло, \\метод Ліса-Александровича

\end{abstract}

\end{document}